\documentclass[submission, Proceedings]{SciPost}

\hypersetup{
    colorlinks,
    linkcolor={red!50!black},
    citecolor={blue!50!black},
    urlcolor={blue!80!black}
}

\usepackage[bitstream-charter]{mathdesign}
\usepackage{bm}
 \usepackage{float}
\usepackage{slashed}
\usepackage{verbatim}
\newcommand{\be}{\begin{eqnarray}}
\newcommand{\ee}{\end{eqnarray}}
\usepackage{leftidx}
\usepackage{multirow}

\DeclareSymbolFont{usualmathcal}{OMS}{cmsy}{m}{n}
\DeclareSymbolFontAlphabet{\mathcal}{usualmathcal}
\renewcommand{\d}{\mathrm{d}}

\newcommand{\xB }{x_{\scriptscriptstyle B}}

\begin{document}

\begin{center}{\Large \textbf{
Sivers asymmetry in inelastic $J/\psi$ leptoproduction at the EIC\\
}}\end{center}

\begin{center}
S. Rajesh\textsuperscript{1$\star$},
U. D'Alesio\textsuperscript{1,2},
A. Mukherjee\textsuperscript{3},
F. Murgia\textsuperscript{1} and
C. Pisano\textsuperscript{1,2}
\end{center}

\begin{center}
{\bf 1} INFN, Sezione di Cagliari, Cittadella Universitaria di Monserrato, I-09042 Monserrato (CA), Italy\\
{\bf 2} Dipartimento di Fisica, Universit\`a di Cagliari, Cittadella Universitaria di Monserrato, I-09042 Monserrato (CA), Italy
\\
{\bf 3} Department of Physics, Indian Institute of Technology Bombay, Mumbai, India
\\
* rajesh.sangem@pg.infn.it\footnote{Now at INFN, Sezione di Perugia, via A. Pascoli snc, 06123,
Perugia, Italy.}
\end{center}

\begin{center}
\today
\end{center}

\definecolor{palegray}{gray}{0.95}
\begin{center}
\colorbox{palegray}{
  \begin{tabular}{rr}
  \begin{minipage}{0.1\textwidth}
    \includegraphics[width=22mm]{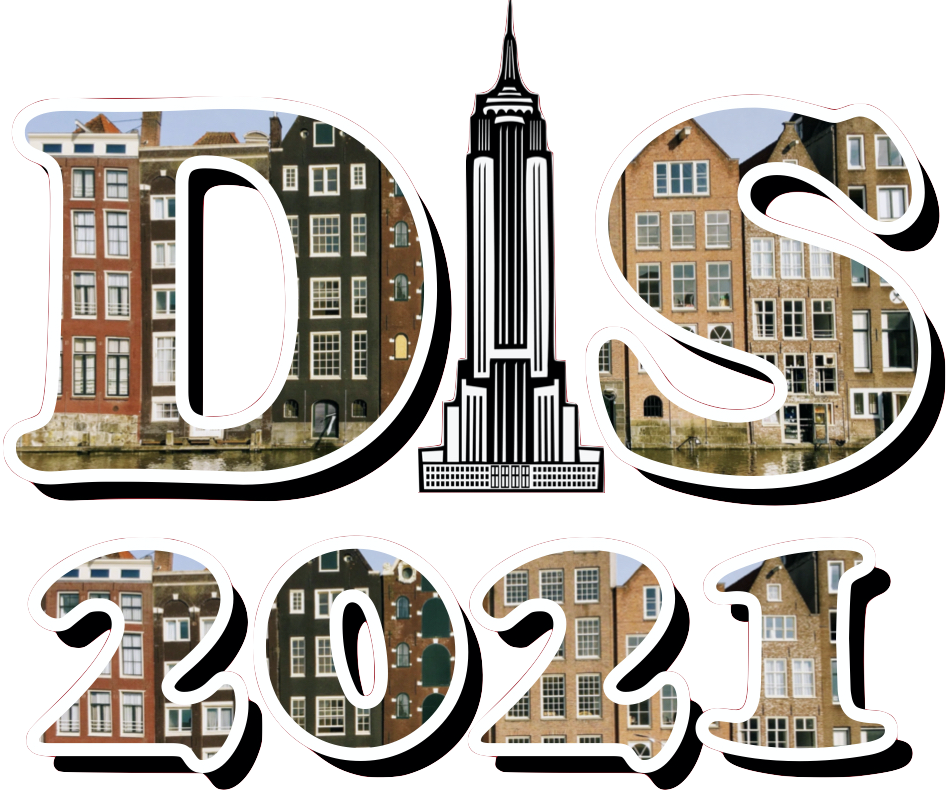}
  \end{minipage}
  &
  \begin{minipage}{0.75\textwidth}
    \begin{center}
    {\it Proceedings for the XXVIII International Workshop\\ on Deep-Inelastic Scattering and
Related Subjects,}\\
    {\it Stony Brook University, New York, USA, 12-16 April 2021} \\
    \doi{10.21468/SciPostPhysProc.?}\\
    \end{center}
  \end{minipage}
\end{tabular}
}
\end{center}

\section*{Abstract}
{\bf
We study the Sivers asymmetry in inelastic $J/\psi$ leptoproduction, $ep^\uparrow \to e+ J/\psi+X$, within a transverse momentum dependent scheme, the so-called generalized parton model (GPM). The effects of final-state interactions are properly taken into account by employing the color-gauge invariant GPM (CGI-GPM). For the $J/\psi$ formation the non-relativistic QCD (NRQCD) framework is adopted. Predictions for unpolarized cross sections and maximized Sivers asymmetries at EIC energies are given.
}

\section{Introduction}\label{sec:intro}
Transverse momentum dependent parton distribution and fragmentation functions, shortly referred to as TMDs, encode information on the 3-dimensional (3D) structure of the proton and are currently playing a growing role in hadron physics. They provide much more insight than the conventional one-dimensional parton distribution functions (PDFs).
Among the eight leading-twist TMDs, the Sivers function~\cite{Sivers:1989cc,Sivers:1990fh} has been receiving special attention both theoretically and experimentally. It has very striking features:  it is time-reversal odd, is related to the partonic orbital angular momentum, which could play a role in the proton spin crisis issue, and is expected to be process dependent. While the quark Sivers function has been extracted in several global fits~\cite{Anselmino:2016uie, Bacchetta:2020gko}, the gluon Sivers function (GSF)~\cite{Mulders:2000sh} is still poorly known, even though some initial attempts have been made~\cite{DAlesio:2015fwo,DAlesio:2019qpk} by fitting mid-rapidity pion and $D$-meson data at RHIC.

$J/\psi$ production has been suggested as a powerful tool to probe the gluon distribution since its discovery. Azimuthal asymmetries for $J/\psi$ production in $ep$ \cite{Mukherjee:2016qxa,Rajesh:2018qks,Kishore:2019fzb} and $pp$ \cite{DAlesio:2019gnu,DAlesio:2020eqo,DAlesio:2019qpk,DAlesio:2017rzj} collisions have been discussed as a way to access the GSF within the generalized parton model (GPM), a phenomenolgical extension of the collinear factorization approach with inclusion of spin and intrinsic transverse momentum effects~\cite{DAlesio:2007bjf}.
In Ref.~\cite{Mukherjee:2016qxa}, the Sivers effect was studied in $ep^\uparrow \to e+ J/\psi+X$ processes at leading order (LO) within the GPM. In such a case the LO subprocess contributes at $z=1$, where $z$ is the energy fraction carried by the $J/\psi$ meson.

In this contribution, we present preliminary results of an ongoing study on the Sivers effect in $ep^\uparrow \to e+ J/\psi+X$ process, within the GPM at next-to-leading order (NLO) accuracy, allowing to probe the GSF in the region $z<1$. Moreover, in order to study the effects of initial- (ISI) and final-state (FSI) interactions on the Sivers asymmetry, we employ the color-gauge invariant GPM (CGI-GPM) approach~\cite{Gamberg:2010tj,DAlesio:2017rzj}. For the $J/\psi$ formation mechanism we will adopt the non-relativistic QCD (NRQCD) effective field theory~\cite{Bodwin:1994jh}, wherein the quarkonium state is described by a double expansion in terms of the strong coupling constant $\alpha_s$ and the relative velocity $v$ of the heavy-quark pair in the quarkonium rest frame. Moreover, the  $c\bar{c}$ pair could be initially produced both in the color singlet (CS) and color octet (CO) states. We will focus on the EIC kinematics, giving predictions both for the unpolarized cross sections and the Sivers asymmetries.

\section{Sivers asymmetry}\label{sec:Sivers}
We consider the (un)polarized electron-proton collision process
\be
e(l)+p^\uparrow(P)\rightarrow e(l^\prime)+ J/\psi (P_h)+X\,,
\ee
where the letters in the round brackets represent the 4-momentum of the corresponding particle, and the arrow in the superscript represents the transverse polarization of the proton.\par
The weighted Sivers asymmetry for $ep^\uparrow \to e+ J/\psi+X$ process is defined as
\begin{equation}\label{eq:AN}
A^{\sin(\phi_h-\phi_S)}_N  \equiv  2\frac{\int \d\phi_S\d\phi_h\sin(\phi_h-\phi_S)
(\d \sigma^\uparrow - \d \sigma^\downarrow)}{\int \d\phi_S\d\phi_h(\d \sigma^\uparrow + \d \sigma^\downarrow)}
 \equiv \frac{\int \d\phi_S \d\phi_h\sin(\phi_h-\phi_S)\d\Delta\sigma(\phi_S,\phi_h)}{\int \d\phi_S\d\phi_h \d\sigma}\,,
\end{equation}
where $\d \sigma^{\uparrow(\downarrow)}=\d\sigma^{\uparrow(\downarrow)}/\d Q^2\, \d y\, \d ^2{\bm P}_T\,\d z$ is the differential cross section with the initial proton polarized along the transverse direction $\uparrow(\downarrow)$ with respect to the lepton plane in the $\gamma^*-p$ center of mass (c.m.) frame. $Q^2=-q^2$ is the virtuality of the photon and $\xB=Q^2/(2P\cdot q)$ is the Bjorken variable. The energy fraction carried by the photon is $y=P\cdot q/P\cdot l$, implying $Q^2=\xB y s$. The inelastic variable is defined as $z=P\cdot P_h/P\cdot q$.
$\phi_S$ and $\phi_h$ are the azimuthal angles of the proton spin and the $J/\psi$ transverse momentum (${\bm P}_T$) respectively.

Assuming TMD factorization within the GPM framework, the unpolarized differential cross section, entering the denominator of Eq.~\eqref{eq:AN}, can be written as
\be\label{unp:eq}
 \frac{\d\sigma}{\d Q^2\, \d y\, \d ^2{\bm P}_T\,\d z}&=&\frac{1}{2s}\frac{2}{(4\pi)^4 z}\sum_{a}\int \frac{\d x_a}{x_a}\, \d^2{\bm k}_{\perp a}\,\delta\left(\hat{s}+\hat{t}+\hat{u}-M^2+Q^2\right)\nonumber\\
 &&{}\times\sum_n\frac{1}{Q^4} f_{a/p}(x_a,k_{\perp a}) L^{\mu\nu}
H^{a,U}_{\mu\nu}[n] \langle 0\mid \mathcal{O}^{J/\psi}(n)\mid 0\rangle\,\,,
\ee
where $a=g,u,\bar{u},d,\bar{d},s,\bar{s}, c, \bar{c}$, $n=\leftidx{^3}{S}{_1^{(1,8)}}, \leftidx{^1}{S}{_0^{(8)}}, \leftidx{^3}{P}{_J^{(8)}}$ with $J=0,1,2$, $s$ is the c.m.~energy squared of the $ep$ process and $M$ the $J/\psi$ invariant mass. In Eq.~\eqref{unp:eq}, $\hat{s}$, $\hat{t}$, $\hat{u}$ are the usual partonic Mandelstam variables.
$f_{a/p}(x_a,k_{\perp a})$ is the unpolarized TMD with light-cone momentum fraction $x_a$ and partonic transverse momentum $k_{\perp a}=|\bm{k}_{\perp a}|$.
The leptonic tensor has the standard form
\be
L^{\mu\nu}&=&8\pi\alpha Q^2\left(-g^{\mu\nu}+\frac{2}{Q^2}\left(l^\mu l^{\prime\nu}+l^\nu l^{\prime\mu}\right)\right)\,,
\ee
where $\alpha$ is the fine structure constant.
$H^{a,U}_{\mu\nu}[n]$ is the squared amplitude of the partonic process $\gamma^\ast+a\rightarrow c\bar{c}[n] +a$, averaged/summed over the spins and colors of the initial/final parton. $H^{a,U}_{\mu\nu}[n]$ is calculated at the perturbative order $\alpha \alpha_s^2$ using NRQCD. The long distance matrix elements (LDMEs), $\langle 0\mid \mathcal{O}^{J/\psi}(n)\mid 0\rangle$, represent the transition probability of the heavy $c\bar{c}[n]$ pair into a $J/\psi$.
The numerator in Eq.~\eqref{eq:AN} is directly sensitive to the Sivers function and within the GPM model reads
\be\label{num:GPM}
\d\Delta\sigma^\mathrm{GPM} &=&\frac{1}{2s}\frac{2}{(4\pi)^4 z}\sum_{a}\int \frac{\d x_a}{x_a}\, \d^2{\bm k}_{\perp a}\,\delta\left(\hat{s}+\hat{t}+\hat{u}-M^2+Q^2\right)\,\sin(\phi_S-\phi_a)\nonumber\\
 &&{}\times\sum_n\frac{1}{Q^4} \, \left(-2 \, \frac{k_{\perp a}}{M_p}\right)\,f_{1T}^{\perp a} (x_a, k_{\perp a})\, L^{\mu\nu}
H^{a,U}_{\mu\nu}[n]\, \langle 0\mid \mathcal{O}^{J/\psi}(n)\mid 0\rangle\,,
\ee
where $f_{1T}^{\perp a}(x_a, {k}_{\perp a })$ is the Sivers function, describing the azimuthal distribution of unpolarized partons in a transversely polarized proton and $\phi_a$ is the azimuthal angle of $\bm{k}_{\perp a}$.

In the CGI-GPM, ISIs and FSIs are taken into account within the single eikonal gluon approximation by following the formalism developed in Ref.~\cite{DAlesio:2017rzj}. Two process-dependent GSFs appear in the formation of a CS state from three colored gluons: the $f$-type and $d$-type gluon Sivers functions, which are respectively related to the antisymmetric and symmetric color combinations. In $ep$ collisions, ISI are absent due to the colorless nature of the virtual photon. Moreover, we find that only the $f$-type GSF is contributing to the Sivers asymmetry, while the $d$-type contribution is zero. This means that quarkonium production in  $ep$ collision is a powerful tool to access the process-dependent $f$-type GSF. Another important result is that the modified color factor associated with the $\leftidx{^3}{S}{_1^{(1)}}$ state is zero in the CGI-GPM approach, which leads to a vanishing Sivers asymmetry in the color-singlet model (CSM). The numerator of the asymmetry in the CGI-GPM framework is then given by
\be\label{num:CGI-GPM}
\d\Delta\sigma^{\mathrm{CGI-GPM}} &=&\frac{1}{2s}\frac{2}{(4\pi)^4 z}\int \frac{\d x_a}{x_a}\, \d^2{\bm k}_{\perp a}\,\delta\left(\hat{s}+\hat{t}+\hat{u}-M^2+Q^2\right)\sin(\phi_S-\phi_a) \nonumber\\
 &&{}\times\left(-2 \, \frac{k_{\perp a}}{M_p}\right)\sum_{n}\frac{1}{Q^4}\,L^{\mu\nu} \Bigg\{\sum_q f_{1T}^{\perp q} (x_a, k_{\perp a})\,H^{q,\mathrm{Inc}}_{\mu\nu}[n]+\nonumber\\
 &&{}\sum_{m=f,d} f_{1T}^{\perp g(m)} (x_a, k_{\perp a})\,
H^{g,\mathrm{Inc}(m)}_{\mu\nu}[n]\Bigg\} \langle 0\mid \mathcal{O}^{J/\psi}(n)\mid 0\rangle\,,
\ee
where $H^{a,\mathrm{Inc}}_{\mu\nu}[n]$ is the perturbative square amplitude calculated by incorporating the FSIs within the CGI-GPM approach, see Ref.~\cite{DAlesio:2017rzj} for more details.
\section{Results}\label{sec:results}
The phenomenological estimates of unpolarized cross sections we are going to show are obtained by adopting a factorized Gaussian ansatz for the TMDs:
\be
f_{a/p}(x, k_\perp) = f_{a/p}(x)\,\frac{e^{-k_\perp^2/\langle k_{\perp a}^2 \rangle}}{\pi\langle k_{\perp a}^2 \rangle }  \,,
\ee
where for quarks we use $\langle k_{\perp q}^2 \rangle= 0.25$ GeV$^2$~\cite{Anselmino:2005nn}, while for gluons $\langle k_{\perp g}^2 \rangle= 1$ GeV$^2$~\cite{DAlesio:2017rzj}. Concerning the LDMEs, we use the BK11 set~\cite{Butenschoen:2011yh}, while the CTEQL1 set~\cite{Pumplin:2002vw} is adopted for the collinear PDFs at the scale $M_T=\sqrt{M^2+P^2_T}$. The QCD partonic subprocesses at order $\alpha\alpha_s^2$ are $\gamma^\ast+g\to J/\psi+g$ and $\gamma^\ast+q(\bar{q})\to J/\psi+q(\bar{q})$, where $q=u,d,s$. Moreover, we consider also the intrinsic charm-quark contribution, i.e.~the $\gamma^\ast +c (\bar{c}) \to J/\psi +c (\bar{c})$ channel. Besides the direct $J/\psi$ production, one has to take into account the feed-down contribution from the decay of excited states and $b$ quarks. We include only the excited state $\psi(2S)$, since the $b$ quark and $\chi_c$ contributions are marginal in the kinematical region under study. The cross section exhibits infrared singularities, particularly for the $ \leftidx{^1}{S}{_0^{(8)}}$ and $\leftidx{^3}{P}{_J^{(8)}}$  states when $z\to 1$. In order to keep the perturbative calculation under control, we consider $0.3<z<0.9$. The lower cut on $z$ is chosen to avoid the resolved $J/\psi$ contribution.
The other kinematical cuts are: $2.5< Q^2 < 100$ GeV$^2$ and $10<W_{\gamma p} < 40$ GeV for the EIC $\sqrt{s}=45$ GeV setup, where $W_{\gamma p}$ is the c.m.~energy of the photon-proton system. For $\sqrt{s}=140$ GeV we use $20<W_{\gamma p} < 80$ GeV.

\begin{figure}[t]
\begin{center}
\includegraphics[width=0.52\textwidth]{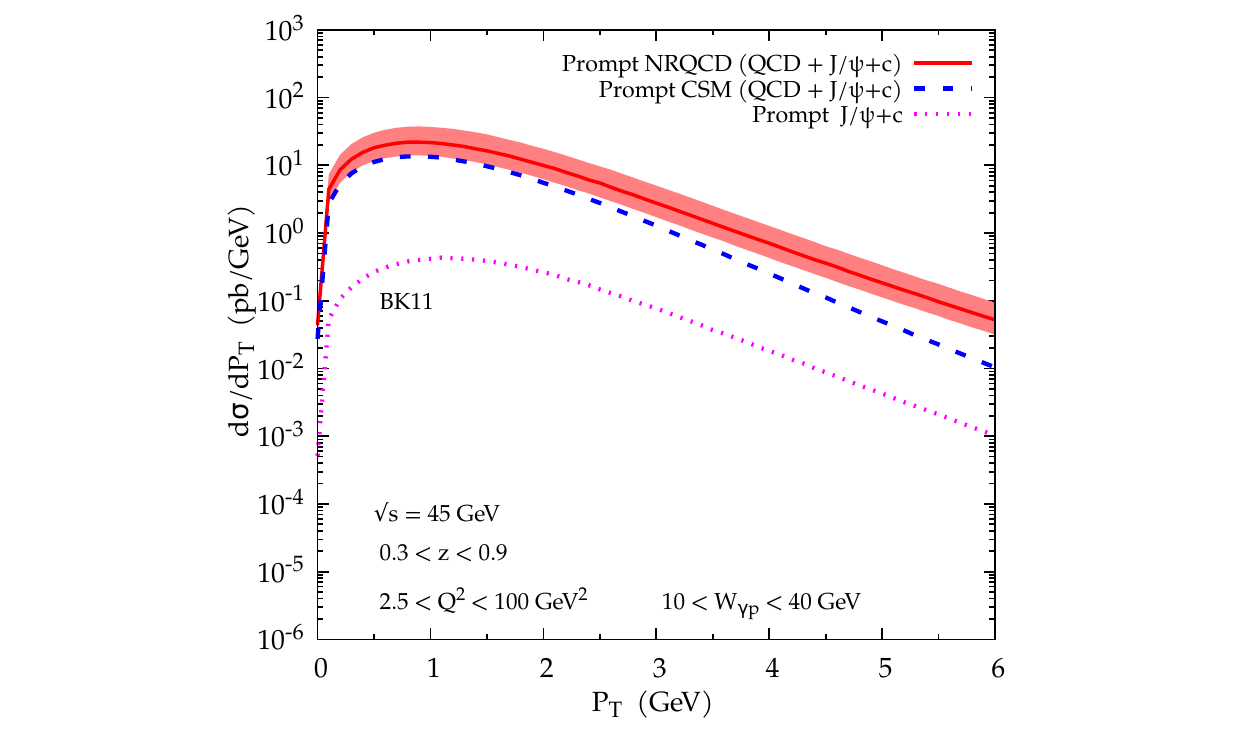}\hspace*{-1cm}
\includegraphics[width=0.52\textwidth]{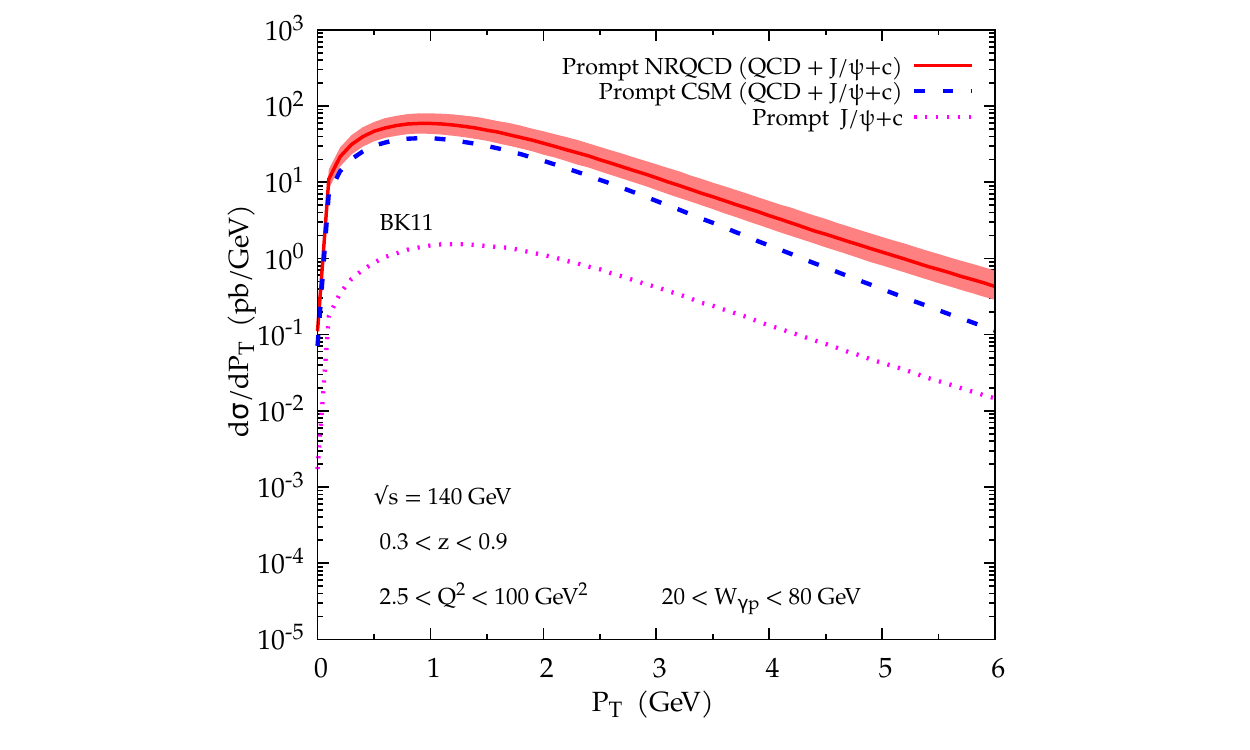}
\caption{Unpolarized differential cross section estimates as a function of $P_T$ for the process
$e p\to e +J/\psi + X$ at $\sqrt s=45$ GeV (left panel) and 140 GeV (right panel), obtained adopting the BK11 LDME set~\cite{Butenschoen:2011yh}: NRQCD (red solid line), CSM (blue dashed line) and intrinsic charm contribution $J/\psi+c$ (magenta dotted line). The band is obtained by varying the factorization scale from $M_T/2$ to $2M_T$.}
\label{fig:unp_EIC}
\end{center}
\end{figure}

In Fig.~\ref{fig:unp_EIC}, estimates of the differential cross section for prompt (direct plus feed-down) $J/\psi$ production as a function of $P_T$  are shown at $\sqrt{s}=45$ GeV (left panel) and $140$ GeV (right panel), for both  NRQCD (red solid curve) and CSM (blue dashed curve). The $J/\psi+c$ curve (magenta dotted line) in Fig.~\ref{fig:unp_EIC} represents the intrinsic-charm contribution, which is almost irrelevant. The band in Fig.~\ref{fig:unp_EIC} is obtained by varying the factorization scale from $M_T/2$ to $2 M_T$.

\begin{figure}[H]
\begin{center}
\includegraphics[width=0.48\textwidth]{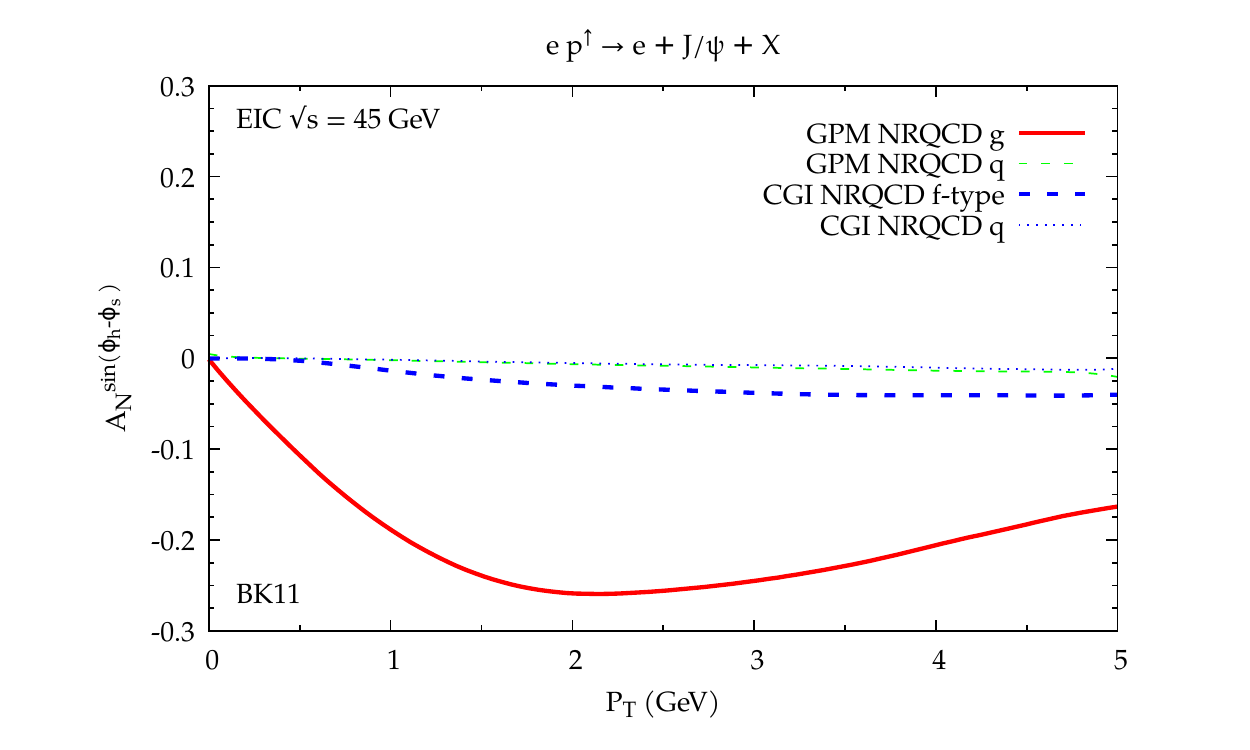}
\includegraphics[width=0.48\textwidth]{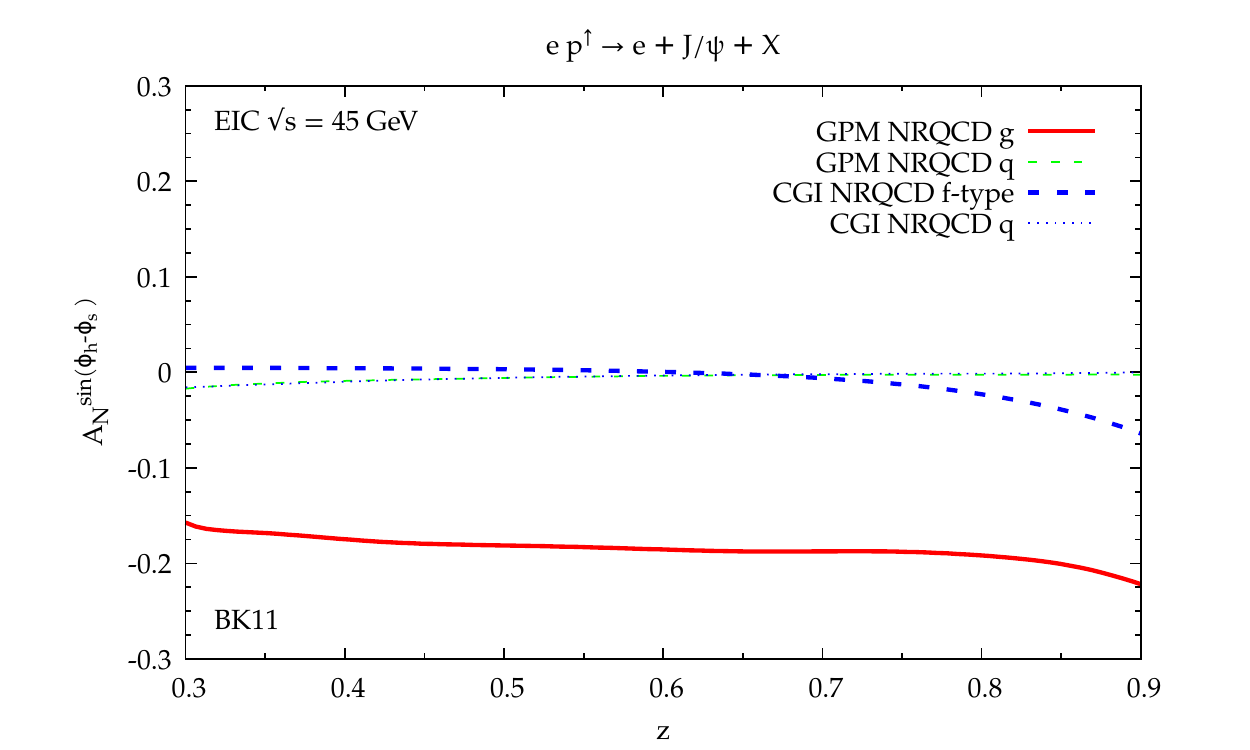}
\caption{Maximized Sivers asymmetry for the $e p^\uparrow\to e +J/\psi + X$ process as a function of $P_T$  (left panel) and $z$ (right panel) obtained with the BK11 LDME set~\cite{Butenschoen:2011yh} at $\sqrt s=45$ GeV: GSF contribution in NRQCD for GPM (red solid thick line), CGI-GPM (blue dashed thick line), quark Sivers contribution in NRQCD for GPM (green dashed thin line) and CGI-GPM (blue dotted thin line).}
\label{EIC45ptz_BK}
\end{center}
\end{figure}

In Fig.~\ref{EIC45ptz_BK} we show the maximized Sivers asymmetry, $A^{\sin(\phi_h-\phi_S)}_N$, as a function of $P_T$ (left panel) and $z$ (right panel) only at $\sqrt{s}=45$ GeV (no strong energy dependence is observed).
Estimates are obtained by saturating the Sivers functions entering Eqs.~\eqref{num:GPM} and  \eqref{num:CGI-GPM}. The asymmetry is mostly dominated by the GSF, while the quark contribution is negligible. This indicates that such observable is a powerful tool to probe the unknown GSF. The GPM predicts negative values around $20\%$. The asymmetry is drastically reduced in size in the CGI-GPM due to color-factor relative cancelations and the absence of the $\leftidx{^3}{S}{_1^{(1)}}$ state contribution and is essentially driven by the $f$-type GSF.

\section{Conclusions}
We have presented a preliminary study of the Sivers asymmetry in $ep^\uparrow \to e+ J/\psi+X$ within the generalized parton model. Final state interactions have been also considered adopting the color-gauge invariant GPM. The NRQCD framework has been used for the $J/\psi$ formation process. Estimates of the unpolarized cross sections for $J/\psi$ production as a function of $P_T$ are given for EIC at $\sqrt s=45$ GeV and 140 GeV energies. The maximized Sivers asymmetry is estimated at $\sqrt s=45$ GeV, showing how it is largely dominated by the gluon Sivers effect both in the GPM and the CGI-GPM, while the quark Sivers contribution turns out to be negligible. Moreover, only the $f$-type GSF contributes to the asymmetry in the CGI-GPM. These findings confirm the potentialities of this analysis and the role of EIC in probing the unknown GSF and its process dependence.

\paragraph{Funding information}
This work is financially supported by Fondazione Sardegna under the project ``Proton tomography at the LHC”, project number F72F20000220007 (University of Cagliari). This project has received funding from the European Union’s Horizon 2020 research and
innovation programme under grant agreement STRONG-2020 - N.~824093.




\bibliography{ref.bib}

\nolinenumbers

\end{document}